# The activated scaling behavior of quantum Griffiths singularity in two-dimensional superconductors

Zihan Cui[1,9], Longxin Pan[1,9], Jingchao Fang[2,9], Shichao Qi[2], Ying Xing[3], Haiwen Liu[4], Yi Liu[1,5,*], Jian Wang[2,6,7,8,*]

[1] Department of Physics, Renmin University of China, Beijing 100872, China

[2] International Center for Quantum Materials, School of Physics, Peking University, Beijing 100871, China

[3] State Key Laboratory of Heavy Oil Processing, College of New Energy and Materials, China University of Petroleum, Beijing 102249, China

[4] Center for Advanced Quantum Studies, Department of Physics, Beijing Normal University, Beijing 100875, China

[5] Beijing Key Laboratory of Opto-electronic Functional Materials & Micro-Nano Devices, Renmin University of China, Beijing 100872, China

[6] Collaborative Innovation Center of Quantum Matter, Beijing 100871, China

[7] CAS Center for Excellence in Topological Quantum Computation, University of Chinese Academy of Sciences, Beijing 100190, China

[8] Beijing Academy of Quantum Information Sciences, Beijing 100193, China

E-mail: jianwangphysics@pku.edu.cn (J.W.) and yiliu@ruc.edu.cn (Y.L.)

[9] These authors contributed equally to this work.

[*] Authors to whom any correspondence should be addressed.

## Abstract

Quantum Griffiths singularity is characterized by the divergence of the dynamical critical exponent with the activated scaling law and has been widely observed in various two-dimensional superconductors. Recently, the direct activated scaling analysis with the irrelevant correction has been proposed and successfully used to analyze the experimental data of crystalline $PdTe_2$ and polycrystalline $\beta$-W films, which provides new evidence of quantum Griffiths singularity. Here we show that the direct activated scaling analysis is applicable to the experimental data in different superconducting films, including tri-layer Ga films and $LaAlO_3/SrTiO_3$ interface superconductor. When taking the irrelevant correction into account, we calculate the corrected sheet resistance at ultralow temperatures. The scaling behavior of the corrected resistance in a comparably large temperature regime and the theoretical fitting of the phase boundary give unambiguous evidence of quantum Griffiths singularity. Compared to the previous method based on the finite size scaling, the direct activated scaling analysis represents a more direct and precise way to analyze the experimental data of quantum Griffiths singularity in diverse two-dimensional superconductors.

## 1. Introduction

As a prototype of quantum phase transition [1,2], the superconductor–insulator or superconductor–metal transition (SIT or SMT) has been studied for decades [3-11], which facilitates the insight of a variety of novel quantum phenomena. Two-dimensional (2D) superconductors represent an ideal platform for the investigation of quantum phase transition owing to the strong quantum fluctuation [2,12]. Previous observations draw our attention to the quantum Griffiths singularity (QGS) of SMT in low-dimensional disordered superconducting systems [13-21]. Contrary to the conventional quantum phase transitions showing a single quantum critical point [3], QGS presents a large transition region defined by the crossing points of magnetoresistance isotherms, which reveals an unusual critical behavior. Moreover, based on the finite size scaling analysis, previous literatures report the divergence of critical exponent $zv$ as the main characteristics of QGS when approaching the infinite-randomness quantum critical point [13], which challenges the prevailing consensus of quantum phase transition in low-dimensional superconductors. This divergent behavior can be attributed to the formation of the spatially separated superconducting islands named as rare regions governed by the effect of quenched disorder [22-26]. The rare regions and the surrounding normal state form a vortex-glass-like phase called quantum Griffiths phase.

Finite-size scaling analysis plays a key role in investigating the critical behavior of quantum phase transition. As for magnetic field induced SMT, the magnetoresistance isotherms cross each other in a single quantum critical point, where the sheet resistance on temperature and magnetic field can be expressed as [1,2]: $R(B,T) = R_c \cdot F((B - B_c)T^{-1/zv})$. Here $F$ is an arbitrary function with $F(0) = 1$, $B_c$ and $R_c$ are the critical magnetic field and critical resistance of the quantum critical point, $z$ and $v$ are the dynamical critical exponent and correlation length exponent, respectively. In the finite size scaling analysis, the sheet resistance as a function of the scaling variable $|B - B_c|t$ at different temperatures can collapse closely into two branches by adjusting the scaling parameter $t$ ($t = (T/T_0)^{-1/zv}$ and $T_0$ is the lowest temperature in the measurement). The value of critical exponent $zv$ can be acquired by the linear fitting between $\ln T$ and $\ln t$, which is normally a constant for conventional SMT [3].

The first evidence of QGS in 2D superconductors was reported in tri-layer crystalline Ga films under perpendicular magnetic field [14]. In contrast to the conventional SMT, the quantum critical regime of QGS shows multiple crossing points of the neighboring magnetoresistance isotherms, which represent the phase boundary between superconducting and normal states at low temperatures. To use the finite size scaling analysis for multiple crossing points in the Ga films, the magnetoresistance isotherms are divided into several groups and each group consists of three to five adjacent $R(B)$ curves. Thus, the crossing region of each group can be roughly regarded as a single crossing point and the finite size scaling analysis is used to yield different values of $zv$. As the hallmark of QGS, the divergence of $zv$ can be well described by the activated scaling law: $zv \propto |B - B_c^*|^{-v\psi}$ with the correlation length exponent $v \approx 1.2$ and tunneling critical exponent $\psi \approx 0.5$ for 2D superconductors [27,28]. Furthermore, the direct activated scaling analysis with a new irrelevant correction was recently developed to quantitatively modify the resistance at low temperatures and describe the phase boundary in ultrathin crystalline $PdTe_2$ film [29] and polycrystalline $\beta$-W film [30]. This analysis is reasonably taken as a direct way to reveal the activated scaling behavior of 2D superconductors and thus demonstrate the existence of QGS [31].

In this paper, the direct activated scaling analysis with the irrelevant correction is successfully used to quantitatively explain the critical behavior as well as the phase boundary of tri-layer Ga film and $LaAlO_3/SrTiO_3$(110) interface superconductor at low temperatures, which gives the direct evidence of QGS. The irrelevant correction of the sheet resistance originates from the finite temperatures and hence decreases with decreasing temperature. Our work indicates that the direct activated scaling analysis with the irrelevant correction can be widely applied to different 2D superconducting systems to reveal the emergence of QGS.

## 2. The direct activated scaling analysis with the irrelevant correction

In the framework of the random transverse field Ising model, the activated scaling form of resistance satisfy [21]:

$$R(B,T) = \Phi\left\{[(B - B_c^*)/B_c^*][\ln(T^*/T)]^{\frac{1}{\nu\psi}}\right\}, \quad (1)$$

where $\Phi$ is unknown arbitrary function, $T^*$ is the characteristic temperature associated with this quantum phase transition, $B_c^*$ is the critical field at zero temperature. In nonzero temperature regime, finite temperature effect can also induce the irrelevant correction and the scaling functional can be written as [21,32]

$$R(B,T) = \Phi\{[(B - B_c^*)/B_c^*][\ln(T^*/T)]^{(1/\nu\psi)}, u[\ln(T^*/T)]^{-y}\}, \quad (2)$$

where $u$ is the leading irrelevant scaling variable and $y > 0$ is the associate irrelevant exponent. Equation (2) can be expanded in the second argument

$$R(B,T) = \Phi_1\left\{[(B - B_c^*)/B_c^*][\ln(T^*/T)]^{\left(\frac{1}{\nu\psi}\right)}\right\}$$
$$+ u[\ln(T^*/T)]^{-y}\Phi_2\{[(B - B_c^*)/B_c^*][\ln(T^*/T)]^{(1/\nu\psi)}\}, \quad (3)$$

Following the previous works [21,29,30], the temperature dependence of the crossing points can be derived from equation (3), which represents the phase boundary of SMT

$$B_c(T) \propto B_c^*\{1 - u[\ln(T^*/T)]^{-(1/\nu\psi)-y}\}. \quad (4)$$

Then we define $x_1(B,T) = [(B - B_c^*)/B_c^*][\ln(T^*/T)]^{(1/\nu\psi)}$, and $x_2(T) = [\ln(T^*/T)]^{-y}$ to simplify the calculation. Equation (3) can be written as $R(x_1, T) = \Phi_1(x_1) + x_2(T) \cdot u\Phi_2(x_1)$. For any fixed $x_1 = x_1^*$ ($x_1^*$ is an arbitrary value of $x_1$), it is easy to find the linear relationship between $R(T)|_{x_1=x_1^*}$ and $x_2(T)$ according to the formula $R(T)|_{x_1=x_1^*} = \Phi_1(x_1^*) + x_2(T) \cdot u\Phi_2(x_1^*)$. Thus, the irrelevant correction $u\Phi_2(x_1^*)$ equals to the slope of $R(T)|_{x_1=x_1^*}$ vs $x_2(T)$ curve, which can be calculated by the linear fitting based on the least square method. After taking out the aforementioned irrelevant correction term, one can obtain the corrected resistance $\tilde{R}(x_1^*)$ (identical to $\Phi_1(x_1^*)$):

$$\tilde{R}(x_1^*) = R(T)|_{x_1=x_1^*} - x_2(T) \cdot u\Phi_2(x_1^*). \quad (5)$$

Theoretically, the $\tilde{R}(x_1)$ curves for different temperatures can collapse into a single curve and this activated scaling behavior provides the direct evidence of QGS.

## 3. Theoretical analysis of QGS in crystalline Ga films and LAO/STO interface

Figure 1 presents the direct activated scaling analysis on 3-ML crystalline Ga film grown by molecular beam epitaxy (MBE). As shown in Fig. 1(a), the magnetoresistance isotherms from 25 to 500 mK of the Ga film cross each other in a comparably large transition region with multiple crossing points, showing the characteristics of QGS. Here the data in Fig. 1(a) are adapted from Ref. [14]. The crossing points of the $R_s(B)$ curves at neighboring temperatures are presented as yellow dots in Fig. 1(b), which represent the phase boundary between the superconducting state and normal state. At low temperatures, the critical field $B_c$ significantly increases with decreasing temperature. To investigate the critical behavior of the Ga film, the phase boundary is theoretically analyzed via the activated scaling method with the irrelevant correction. As presented in Fig. 1(b), the experimental phase boundary at low temperatures from 30 mK to 234 mK can be well fitted by equation (4), consistent with our theoretical model. Furthermore, we calculated the corrected resistance $\tilde{R}$ as a function of $x_1$ by equation (5) to consider the irrelevant correction, where $x_1(B,T) = [(B - B_c^*)/B_c^*][\ln(T^*/T)]^{(1/\nu\psi)}$ and $\nu\psi \approx 0.6$. In Fig. 1(c), the $\tilde{R}(x_1)$ curves from 25 mK to 198 mK collapse into upper and lower two branches, revealing the activated scaling behavior of QGS. The best fitting in Fig. 1(b) and Fig. 1(c) yields the fitting parameters $T^* = 7.0$ K, $B_c^* = 2.632$ T, $y = 3.08$, $u = 20.3$, which are summarized in Table 1. The theoretical fitting of the phase boundary in Fig. 1(b) and the activated scaling behavior in Fig. 1(c) give the direct and unambiguous evidence of QGS in the Ga film.

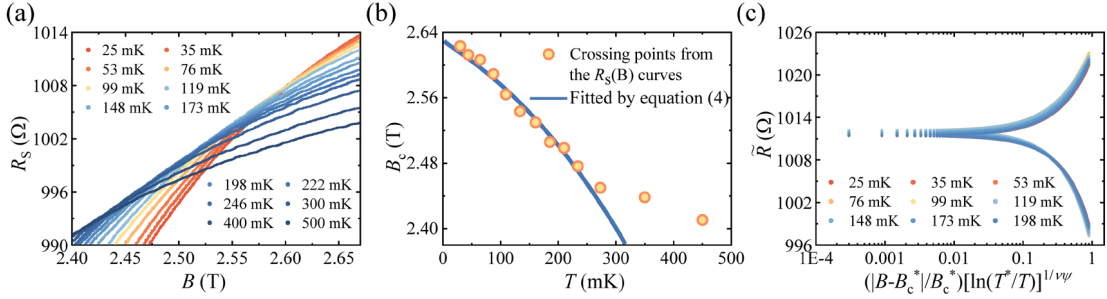

**Figure 1.** The QGS in the 3-ML crystalline Ga film under perpendicular magnetic field. (a) The magnetic field dependence of sheet resistance $R_s$ at various temperatures ranging from 25 mK to 500 mK. The data is adapted from Ref. [14]. (b) Crossing points from the magnetoresistance isotherms at neighboring temperatures. The blue solid line is the fitting curve (equation (4)) from the activated scaling analysis with the irrelevant correction. (c) The direct activated scaling analysis of the corrected resistance $\tilde{R}$ curves from 25 mK to 198 mK considering the irrelevant correction. Here $\tilde{R}$ is calculated by equation (5).

Importantly, the irrelevant correction plays a crucial role in the direct activated scaling analysis on QGS. As shown in Fig. 2, we calculate the irrelevant correction $\Delta R$ at various temperatures from 25 mK to 198 mK and magnetic fields from 2.46 T to 2.76 T based on the following equation. According to equation (3), the irrelevant correction reads

$$\Delta R = u[\ln(T^*/T)]^{-y} \Phi_2\{[(B - B_c^*)/B_c^*][\ln(T^*/T)]^{(1/\nu\psi)}\}. \tag{6}$$

As shown in Fig. 2, $\Delta R$ is below 10 $\Omega$ in this regime, much smaller than the measured sheet resistance $R_s$ around 1000 $\Omega$. Note that $\Delta R$ decreases with decreasing temperature, consistent with our theoretical model that the irrelevant correction is predicted to vanish at the zero-temperature limit.

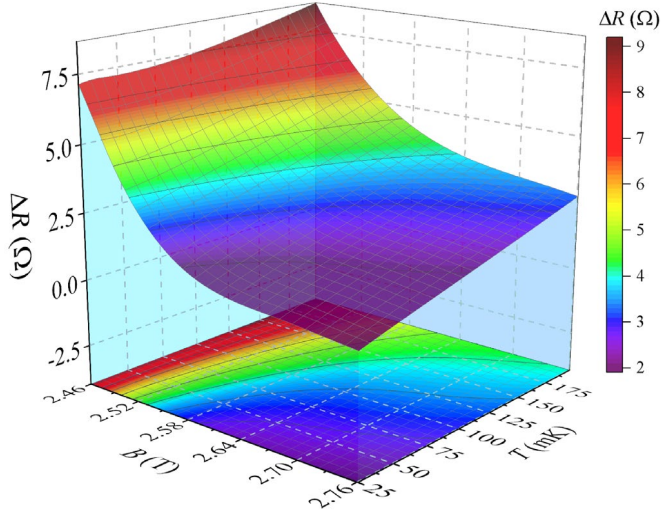

**Figure 2.** The magnetic field and temperature dependence of the irrelevant correction $\Delta R$ for the 3-ML Ga film. The color represents the value of $\Delta R$. The contours in gray solid lines are guides to the eye.

The direct activated scaling analysis with the irrelevant correction successfully explains the feature of QGS in the ultrathin crystalline Ga film. To show whether this method can be applicable in various 2D superconductors, more theoretical analysis on the interface superconductor LaAlO$_3$/SrTiO$_3$ (LAO/STO) is carried out. Figure 3(a) shows the magnetoresistance of different temperatures from 20 mK to 175 mK for the LAO/STO system and the large transition region is the signature of QGS. Here the experimental data in Fig. 3(a) are adapted from Ref. [16]. In Fig. 3(b), the crossing points of $R_s(B)$ curves are shown as yellow dots, revealing the phase boundary of SMT. Furthermore, the phase boundary below 160 mK is theoretically fitted by the activated scaling model with the irrelevant correction (equation (4)). The corrected resistance $\tilde{R}(x_1)$ curves well collapse with each other in Fig. 3(c), which directly demonstrates the existence of QGS in LAO/STO superconductor. The best fitting of the phase boundary and the activated scaling analysis yields the parameters of $T^* = 1.0\text{ K}$, $B_c^* = 0.430\text{ T}$, $y = 2.10$, $u = 1.86$. Furthermore, in Fig. 3(d), the irrelevant correction $\Delta R$ of LAO/STO decreases with decreasing temperature, showing similar temperature dependence of that in the Ga film.

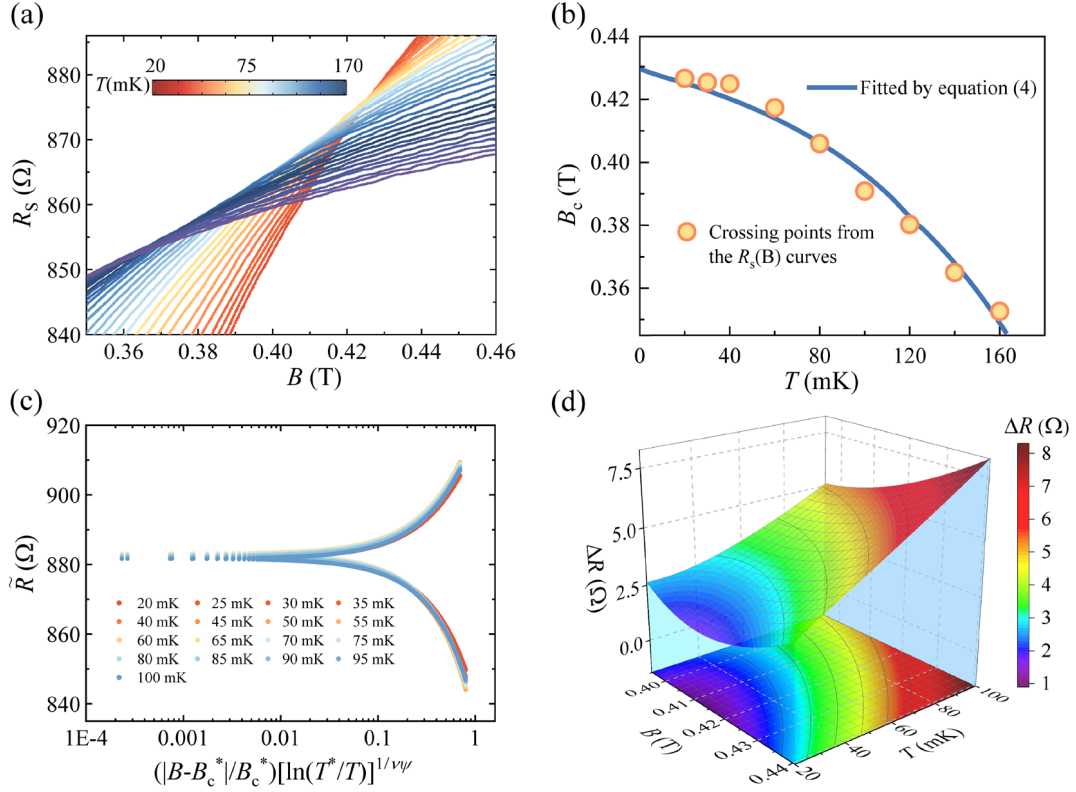

**Figure 3.** The QGS in LAO/STO interface superconductor under perpendicular magnetic field. (a) The magnetic field dependence of $R_s$ at various temperatures ranging from 20 mK to 170 mK, adapted from Ref. [16]. (b) Crossing points from the neighboring magnetoresistance isotherms. The blue solid line is the theoretical fitting curve (equation (4)). (c) The direct activated scaling analysis of the corrected resistance $\tilde{R}$ curves from 20 mK to 100 mK when taking the irrelevant correction into account. (d) The magnetic field and temperature dependence of the irrelevant correction $\Delta R$ for the LAO/STO interface superconductor. The color represents the value of $\Delta R$. The contours in gray solid lines are guides to the eye.

The fitting parameters of the direct activated scaling analysis on QGS for 2D crystalline superconductors, including Ga film, LAO/STO and PdTe$_2$ film [29], are summarized into Table 1. For 2D superconductors, the critical exponent $\nu\psi$ is a constant of 0.6, which is fixed in our theoretical fitting. $B_c^*$ is the critical field of the quantum critical point at zero temperature, which is related to the specific superconducting systems and the orientation of magnetic field. The parameter $T^*$ is the characteristic temperature associated with this quantum phase transition [33], and $\ln(T^*/T)$ is the effective length scale [30]. Normally, $T^*$ is larger for the 2D superconducting systems with larger superconducting transition temperature. The leading irrelevant scaling variable $u$ and the associate irrelevant exponent $y$ are defined by the renormalization group theory [34]. As shown in Table 1, $y$ is on the order of 1 for different 2D superconductors, which can satisfy the requirement of the activated scaling model that $y$ should be larger than 0.

Table 1. The fitting parameters of the activated scaling analysis on QGS.

|  | $T^*$(K) | $B_c^*$(T) | $y$ | $u$ |
|---|---|---|---|---|
| 3-ML Ga film under $B_\perp$ | 7.0 | 2.632 | 3.08 | 20.3 |
| LAO/STO under $B_\perp$ | 1.0 | 0.430 | 2.10 | 1.86 |
| 4-ML PdTe$_2$ film under $B_\perp$ | 1.6 | 0.980 | 1.56 | 0.633 |
| 4-ML PdTe$_2$ film under $B_\parallel$ | 2.0 | 15.28 | 1.62 | 0.308 |

## 4. Conclusions

In conclusion, our work demonstrates that the direct activated scaling analysis with the irrelevant correction can successfully explain the main characteristics of QGS in 3-ML crystalline Ga film and LAO/STO interface superconductors. The activated scaling behavior of the corrected resistance and the theoretical fitting of the phase boundary directly demonstrate the emergence of QGS in 2D superconductors.

Moreover, a previous work reports the observation of anomalous QGS in the ultrathin crystalline Pb films [20], where the perpendicular critical field decreases with decreasing temperatures in the low temperature regime. This anomalous phase boundary in the Pb films is ascribed to the superconducting fluctuation with strong spin-orbit coupling and can not be fitted by the theoretical model in this work. Thus, the direct activated scaling analysis needs to be improved to explain the characteristics of anomalous QGS. Our work will stimulate further studies on the analysis and understanding of novel quantum phase transitions in 2D superconductors.

## Data availability statement

The data that support the findings of this study are available upon reasonable request from the authors.

## Acknowledgements

This work was financially supported by the National Key Research and Development Program of China (Grant No. 2018YFA0305604), the National Natural Science Foundation of China (Grant No. 11888101, No. 12174442, No. 11974430), Beijing Natural Science Foundation (Z180010), the Strategic Priority Research Program of Chinese Academy of Sciences (Grant No. XDB28000000), the Fundamental Research Funds for the Central Universities and the Research Funds of Renmin University of China.